# Higher-order topological spin Hall effect of sound*

Zhi-Kang Lin(林志康)[1], Shi-Qiao Wu(吴世巧)[1], Hai-Xiao Wang(王海啸)[2†], and Jian-Hua Jiang(蒋建华)[1‡]

[1]School of Physical Science and Technology, and Collaborative Innovation Center of Suzhou Nano Science and Technology, Soochow University, 1 Shizi Street, Suzhou, 215006, China

[2]School of Physical Science and Technology, Guangxi Normal University, Guilin, 541004, China

*Supported by the National Natural Science Foundation of China under Grant No 11675116, 11904060, and the Jiangsu distinguished professor funding and a project funded by the Priority Academic Program Development of Jiangsu Higher Education Institutions (PAPD).

**To whom correspondence should be addressed. Email: hxwang@gxnu.edu.cn

jianhuajiang@suda.edu.cn

**Abstract** We propose theoretically a reconfigurable two-dimensional (2D) hexagonal sonic crystal with higher-order topology protected by the six-fold, $C_6$, rotation symmetry. The acoustic band gap and band topology can be controlled by rotating the triangular scatterers in each unit-cell. In the nontrivial phase, the sonic crystal realizes the topological spin Hall effect in a higher-order fashion: (i) The edge states emerging in the bulk band gap exhibits partial spin-momentum locking and are gapped due to the reduced spatial symmetry at the edges. (ii) The gapped edge states, on the other hand, stabilize the topological corner states emerging in the edge band gap. The partial spin-momentum locking is manifested as pseudo-spin-polarization of edge states away from the time-reversal invariant momenta, where the pseudospin is emulated by the acoustic orbital angular momentum. We reveal the underlying topological mechanism using a corner topological index based on the symmetry representation of the acoustic Bloch bands.

---

*Introduction.*---Topological phases and topological phase transitions in both fermionic and bosonic systems have attracted tremendous research interest in the past decades [1-4]. The concept of topological crystalline insulators (TCIs) [5], where band topology is protected by the crystalline symmetry, makes the physical realizations of topological bands feasible in classical wave systems [6-28]. Nevertheless, the reduction of spatial symmetry on the edge boundaries of TCIs often gap the edge states, resulting in less robust transport on the edges. On the other hand, though the edge states may open spectral gaps, the corner boundary of the edges may support topologically-protected, nontrivial corner states, yielding the concept of higher-order topological insulators (HOTIs) [29-53]. Differing from the strong topological insulators, which obey the conventional bulk-edge correspondence, a $l$-th order topological insulator in $n$-dimension has topological boundary states on the $n - l$ dimensional boundaries. For instance, in two-dimensional (2D) systems, second-order topological insulators exhibit one-dimensional (1D) gapped edge states and zero-dimensional (0D) in-gap corner states. The corner states emerge as the topological boundary states of the edge boundary and is rigorously connected to the bulk topology through the bulk-corner correspondence. Here, as a special conceptual quest, our aim is to examine whether the acoustic analog of quantum spin Hall insulators exhibit higher-order topological phenomena.

In this letter, we revisit a 2D hexagonal sonic crystal (SC) comprised of triangular rods with the $C_6$ symmetry [26], where topological transitions can be triggered by rotating the triangular scatterers. As depicted in Fig. 1(a), the 2D SC (with lattice constant $a$) consists of a honeycomb-lattice array of triangular scatterers placed in the air. The scatterers, which are made of epoxy rods, has the shape of a regular triangular with the side length $l$. With the $C_6$ symmetry, the orientation of the triangular scatterers is characterized by a single geometry parameter, the rotation angle $\theta$ (ranging from $-60°$ to $60°$ with a periodicity of $120°$). The rotation angle enables the configurability of the acoustic bands and the band topology, as shown below.

*Acoustic analog of quantum spin Hall insulator.*---At the heart of the acoustic analog of quantum spin Hall insulator is the parity inversion at the time-reversal invariant momenta, as inspired by the Bernevig-Hughes-Zhang (BHZ) model for the quantum spin Hall insulators [1-2]. To illustrate the band inversion, the acoustic bands in SCs with $\theta = 0°$ and $60°$ are shown in Figs. 1(b) and 1(c), respectively. The acoustic band structures are calculated numerically using the commercial finite-element solver, COMSOL MULTIPHYISCS. The symmetry-eigenvalues at the high-symmetry points (HSPs) of the Brillouin zone are analyzed using the phase profiles of the pressure fields, Arg $(P)$ ($P$ denotes the acoustic pressure, which acts as the wavefunction in acoustics), of the acoustic Bloch functions for both cases in Figs. 1(b) and 1(c). We notice that the symmetry eigenvalues for these two cases are identical at the HSPs except the $\Gamma$ point. At the $\Gamma$ point, the band inversion between the dipolar doublet states (i.e., $p_x$- and $p_y$-like states) and the quadrupolar doublet states (i.e., $d_{x^2-y^2}$- and $d_{xy}$- like states) is the key for the acoustic analog of quantum spin Hall insulator. The degenerate dipolar (quadrupolar) states can form a new representation with finite orbital angular momenta (OAM), i.e., $p_\pm = p_x \pm ip_y$ ($d_\pm = d_{x^2-y^2} \pm id_{xy}$). The acoustic OAM emulate the pseudospins and enable an analog with the BHZ model within the framework of the $k \cdot p$ theory around the $\Gamma$ point for the acoustic bands (see Appendix for details).

Depending on the symmetry representation at the $\Gamma$ point, the acoustic band gap between the third and fourth bands can be trivial or topological. For the SC with $\theta = 0°$, the $p$-like states are below the $d$-like states, which corresponds to the "normal insulator" (NI) phase. In contrast, for the SC with $\theta = 60°$, the parity inversion at the $\Gamma$ point gives a "topological crystalline insulator" (TCI) protected by the $C_6$ symmetry. As shown in the Appendix, the TCI (NI) has a $k \cdot p$ Hamiltonian similar to the BHZ Hamiltonian with negative (positive) Dirac mass, i.e., the Hamiltonian for the quantum spin Hall insulator (normal insulator).

To visualize the topological transition, we plot the evolutions of the frequencies of the dipolar and quadrupolar modes at the $\Gamma$ point as functions of the rotation angle in Fig. 1(d). It is seen that the band gap experiences processes of closing and reopening by rotating the triangular rods, indicating topological transitions associated with the band inversions. The reconfigurable SC investigated in this work fits well for the study of higher-order topological phenomena.

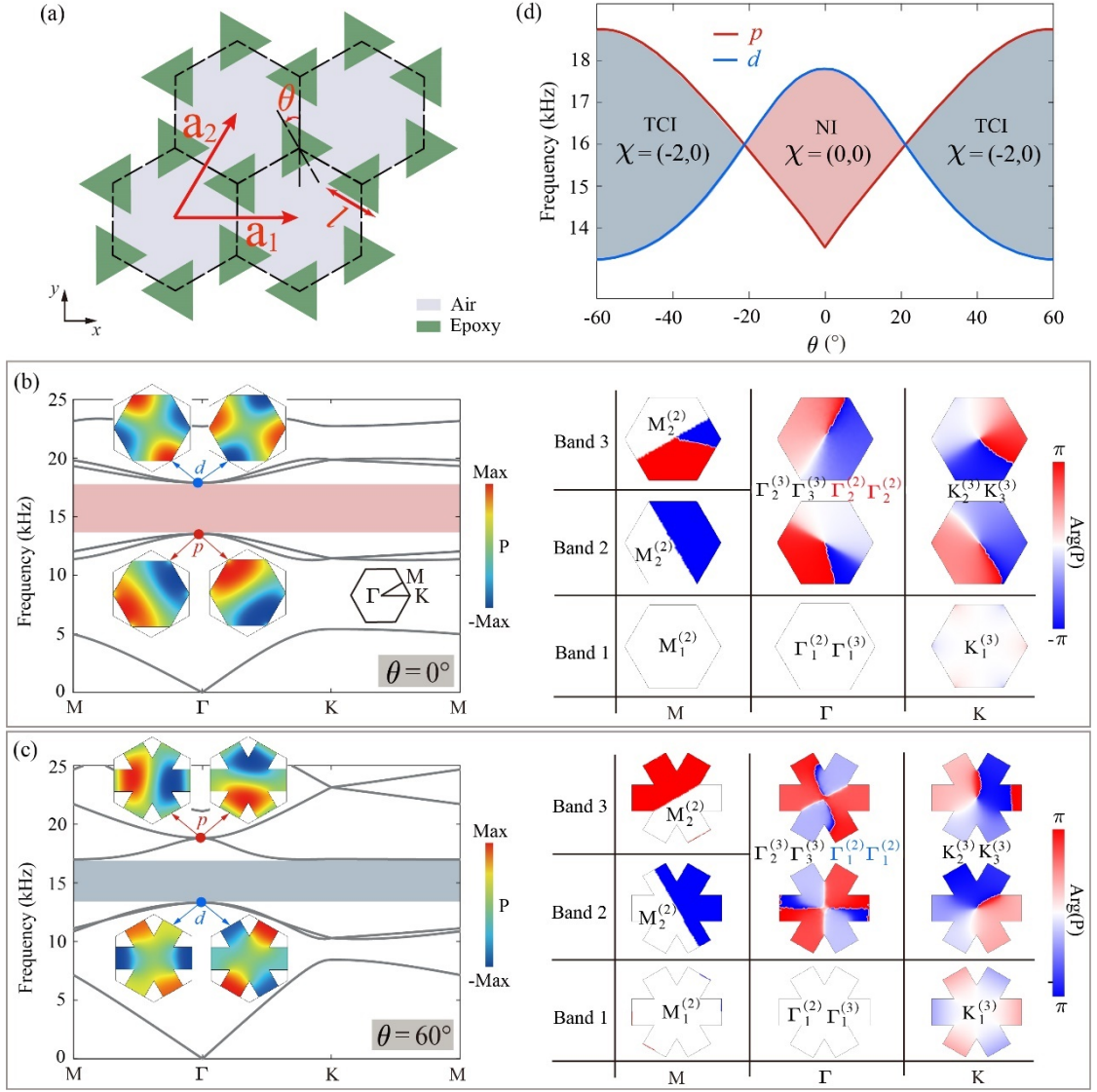

**Figure 1** (a) Schematic of the 2D SC consisting of a honeycomb array of triangular scatterers. The two basis vectors of the hexagonal lattice are labeled as $\mathbf{a_1}$ and $\mathbf{a_2}$, $l$ and $\theta$ refer to the side length of the triangles and the rotation angle, respectively. (b, c) The acoustic band structure of (b) the acoustic normal insulator phase with $\theta = 0°$ and (c) the acoustic topological crystalline insulator phase with $\theta = 60°$. Inset: the eigenstates of the $d$-doublet states and the $p$-doublet states. Right panel: The phase profiles of the pressure fields, $Arg(P)$, of the acoustic Bloch functions below the band gap at the HSPs. The rotation eigenvalues are also labeled. (d) The topological phase diagram manifested as the evolution of the frequencies of the $d$-doublet and $p$-doublet at the $\Gamma$ point versus the rotation angle.

*Gapped edge states.*---When the TCI and NI SCs are placed together, edge states emerge in the bulk band gap. These edge states, as shown in Fig. 2(a), exhibit partial spin-momentum locking. The partial spin-momentum locking is due to the finite spectral gap in the edge states, which is similar to the gapped edge states in the magnetized quantum spin Hall insulators [54]. While the latter is due to time-reversal symmetry breaking, the gapped edge states here originate from the breaking of the $C_6$ symmetry at the edges. Since no edges can preserve the $C_6$ symmetry, the gap opening in the edge states are inevitable.

The edge gap opening induces mixing between the edge states with opposite pseudospins, as revealed initially in Ref. [7]. As a consequence, the edge states become partially spin-moment locked: away from the time-reversal invariant momenta (i.e., $k_x = 0, \frac{\pi}{a}$), the pseudospin polarization is still prominent, while close to the time-reversal invariant momenta the pseudospin polarization is suppressed. At the time-reversal invariant momenta, the pseudospin polarization (i.e., the OAM) strictly vanishes due to time-reversal symmetry. To elucidate the acoustic pseudospin, we present the phase and amplitude profiles of the acoustic pressure field for the pseudospin-polarized edge states (A and B) in Fig. 2(b). The OAM of the edge states are manifested in two aspects. First, the vortices in the phase profiles indicate finite OAM. Second, the distributions of the acoustic energy flows (the green arrows) indicate the winding of the acoustic momentum and the finite OAM.

To demonstrate the partial pseudospin-momentum locking, we use a source with OAM to excite the pseudospin-polarized edge states. The source is composed of three point sources with a phase delay $\pm\frac{2\pi}{3}$ between them. Depending on whether the phase winding is clockwise or anti-clockwise, acoustic waves with OAM $\pm 1$ can be excited.

Although the OAM of the edge states are not quantized, in the case without pseudospin mixing, the source with OAM $1$ can excite only the edge states with positive OAM, while the source with OAM $-1$ can excite only the edge states with negative OAM. These two types of edge states propagate in opposite directions, leading to OAM-selective excitation and unidirectional edge states propagations. Such phenomena have been shown experimentally in Refs. [14,18]. While in these studies the pseudospin mixing is ignored, we show here that the pseudospin mixing induces significant reduction of OAM-selective excitation.

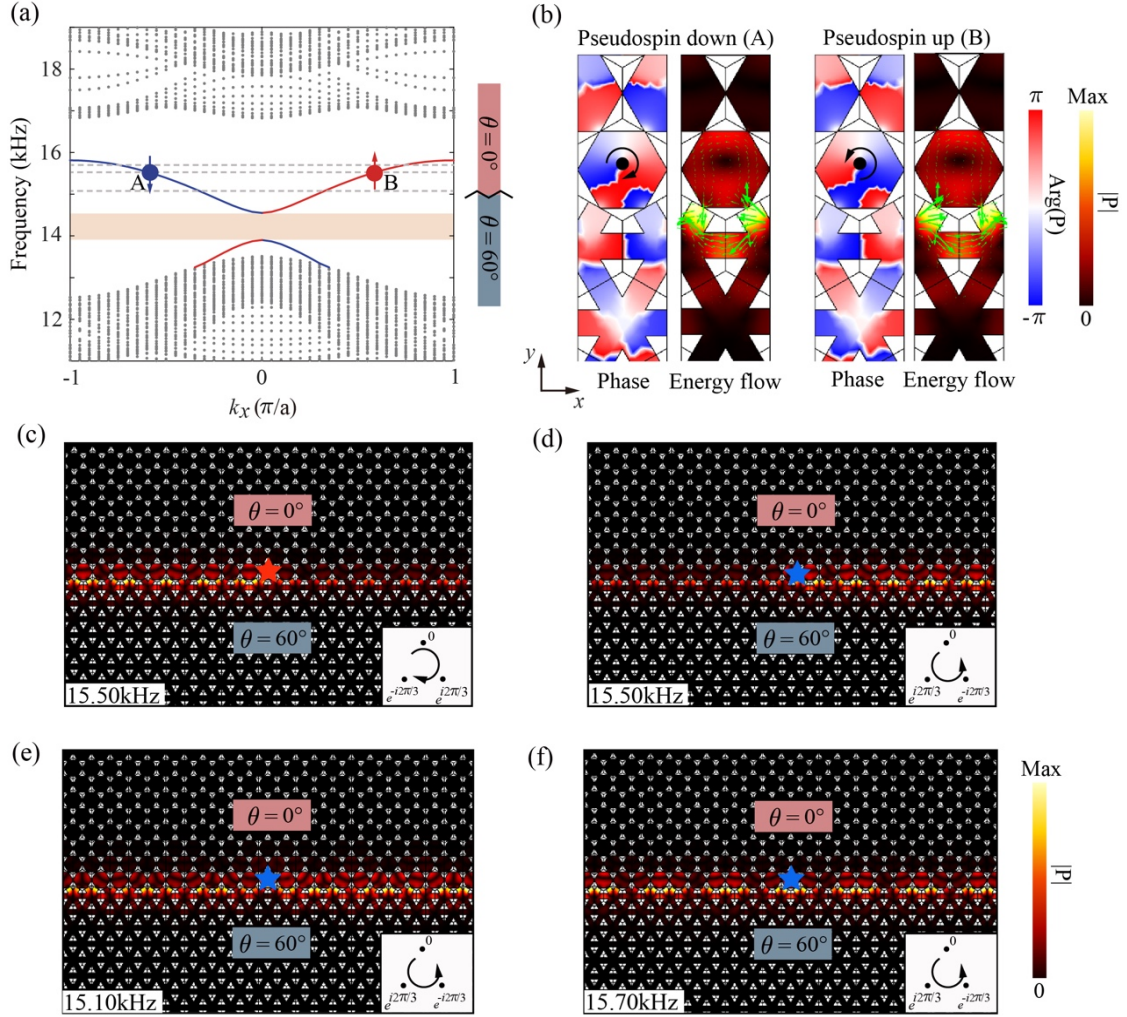

**Figure 2** (a) Left panel: calculated acoustic band structure of the edge boundary between the SC with $\theta = 0°$ (NI) and the SC with $\theta = 60°$ (TCI) for the zigzag edge along the $x$ direction. Right panel: schematic of the ribbon-shaped supercell consisting of the NI and the TCI. (b) Phase and amplitude of the acoustic pressure fields for the edge state A with pseudospin down and the edge state B with pseudospin up [depicted in figure (a); their frequency is 15.5 kHz]. The black arrows in the phase profiles, $\mathrm{Arg}(P)$, indicate the winding directions of the phase vortices (the black dots indicate the vortex centers). The green arrows indicate the distributions of the energy flow for the edge states. (c) and (d): The simulation of the OAM-selective excitation of the edge states at frequency 15.5 kHz. (e) and (f): Suppressed OAM-selectivity in the excitation of the edge states simulated at 15.1 and 15.7 kHz, respectively.

Figs. 2(c) and 2(d) show the partial OAM-selective excitations at 15.5 kHz for the edge states. The three point sources, forming a regular triangle, are placed in the hexagonal region right above the zigzag boundary. The center of the three point sources approximately coincides with the center of the phase vortices in Fig. 2(b). It is seen that the OAM of the source still determines the propagating direction of the majority acoustic energy. Reverse the OAM of the source will switch such a propagating direction. However, at higher or lower frequency, 15.7 or 15.1 kHz, such OAM-selective excitation is strongly suppressed, which is mainly due to the pseudospin-mixing in the edge states. Such mixing is strongly enhanced near the time-reversal invariant momenta of the edge states (i.e., $k_x = 0, \frac{\pi}{a}$).

*Higher-order topology.*---To reveal the higher-order topology in the TCI, we utilize the symmetry-eigenvalues of the acoustic Bloch functions at the HSPs to deduce the bulk topological index, using the method elaborated in Ref. [53]. For an HSP denoted by the symbol $\Pi$, the $C_n$-rotation eigenvalue of the Bloch functions can only be $\Pi_p^{(n)} = e^{2\pi i(p-1)/n}$ with $p = 1, \dots, n$. The HSPs of the Brillouin zone for 2D hexagonal lattices are $\Gamma, M, K$. The full set of the rotation eigenvalues at the HSPs is redundant due to the time-reversal symmetry and the conservation of the number of bands below the band gap. The minimum set of indices that characterize the band topology can be obtained by using the following quantities:

$$\left[\Pi_p^{(n)}\right] = \#\Pi_p^{(n)} - \#\Gamma_p^{(n)}, \tag{1}$$

where $\#\Pi_p^{(n)}$ and $\#\Gamma_p^{(n)}$ are the number of bands below the band gap with eigenvalues $\Pi_p^{(n)}$ and $\Gamma_p^{(n)}$, respectively, with $\Pi = M, K$. In this scheme, the symmetry eigenvalues at the $\Gamma$ point are taken as the reference. The resulting topological classification of insulators with the $C_6$ symmetry is given by the indices $\chi$ as follows

$$\chi = \left(\left[M_1^{(2)}\right], \left[K_1^{(3)}\right]\right). \tag{2}$$

From the phase profiles at the HSPs below the band gap, one can directly obtain the symmetry eigenvalues [see the right panels of Figs. 1(b) and 1(c)]. The topological indices are $\chi = (0,0)$ for the case with $\theta = 0°$ (NI), and $\chi = (-2,0)$ for the case with $\theta = 60°$ (TCI). The NI indeed has a trivial index, while the TCI has nontrivial index due to the parity inversion at the $\Gamma$ point. It should be emphasized that insulators with distinct indices belong to different topological classes and cannot be deformed into one another except closing the bulk energy gap or breaking the $C_6$ symmetry.

According to Ref. [53], the bulk topological invariants in Eq. (2) are connected to the bulk-induced corner topological index, $Q_c$, as

$$Q_c = \left(\frac{1}{4}\left[M_1^{(2)}\right] + \frac{1}{6}\left[K_1^{(3)}\right]\right) \bmod 1. \tag{3}$$

From the bulk topological indices $\chi = \left(\left[M_1^{(2)}\right], \left[K_1^{(3)}\right]\right)$, we find that for the NI, the corner index is $Q_c = 0$, whereas for the TCI, the corner index is $Q_c = \frac{1}{2}$. The difference between them is $\Delta Q_c = \frac{1}{2}$. Such a nontrivial corner index difference indicates the emergence of the topological corner states at the corner boundaries between the NI and TCI SCs, as schematically shown in Fig. 3(a). The switch of the corner index associated with the bulk topological transitions as a function of the rotation angle $\theta$ is shown in Fig. 3(b).

*Topological corner states.*---To unveil the key manifestation of the higher-order topology in the acoustic TCI, we show the eigenstates spectrum of a hexagonal-shape large structure where the SC with $\theta = 60°$ is surrounded by the SC with $\theta = 0°$ in Fig. 3(c). The structure, as depicted in the inset, has both edge and corner boundaries where the edge and corner states reside, respectively. The eigenstates thus include the bulk (black dots), edge (orange dots) and corner (blue dots) states. There are six degenerate corner states emerging in the common spectral gap of the edge and bulk. Each corner has one corner state. The local densities of states in the bulk, edge and corner regions are presented in the right panel of Fig. 3(c). For the sake of graphic presentations, a finite Lorentzian broadening of 10Hz is used in the calculation of the density of states. The local densities of states also show the emergence of the topological corner states in the common spectral gap of the edge and bulk.

In Fig. 3(d), we plot the summation of the intensity of the acoustic wavefunctions of the six corner states, which exhibits $C_6$-symmetric patterns. Due to the finite-size effect, these corner states interact and hybridize with each other. Such interactions also lead to slight splitting between the degenerate corner states, as seen in Fig. 3(c). The wavefunctions of the hybridized corner states show exhibit interesting features. As shown in Figs. 3(e) and 3(f), the corner wavefunctions at 14.17 and 14.18 kHz exhibit opposite OAM (as indicated by the distributions of the acoustic energy flow) at the nearby corners. The net OAM for both corner states vanishes according to the time-reversal symmetry.

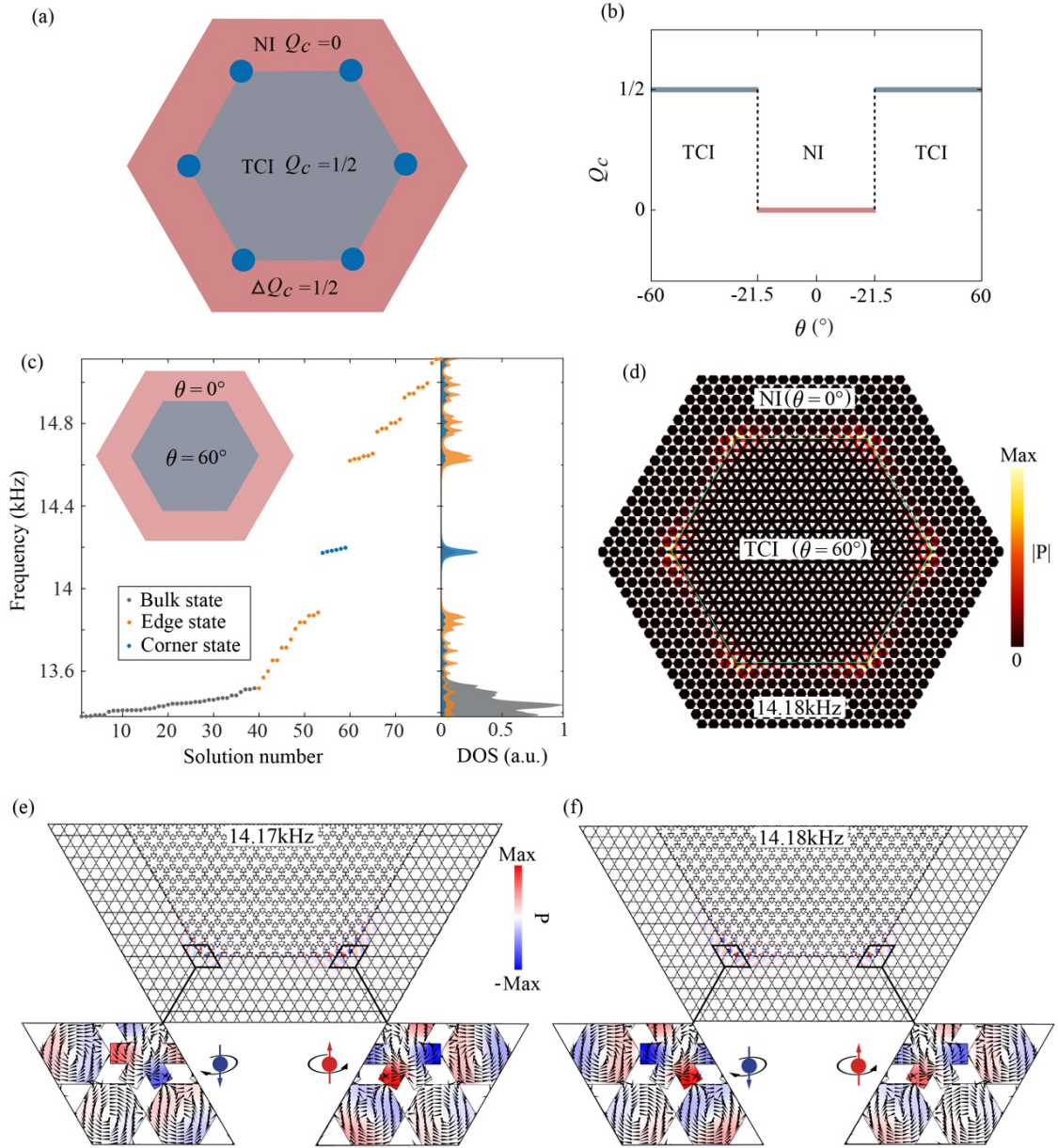

**Figure 3** (a) Schematic of the bulk-corner correspondence in a large hexagonal structure. The corner states (labeled by the blue dots) are dictated by the bulk-induced corner index $Q_c$. (b) Evolution of the corner index $Q_c$ as a function of the rotation angle. (c) Calculated eigen-spectrum (left panel) and local density of states (right panel) for the bulk, edge and corner in a large but finite structure (depicted in the inset) with both edge and corner boundaries between the SCs with $\theta = 60°$ (TCI) and $\theta = 0°$ (NI). (d) The amplitude profile of the acoustic pressure field for the corner states. (e) and (f) The zoom-in distributions of the acoustic pressure field (color) and the energy flow (black arrows) for the corner states at 14.17 and 14.18 kHz.

*Conclusion and outlook.*---In summary, we demonstrate the higher-order topological spin Hall effect of sound in a reconfigurable 2D hexagonal SC. The acoustic band gap and band topology can be controlled by the rotation angle of the triangular scatterers in each unit-cell. The topological phase transitions are well-described by the picture of parity inversion. In the nontrivial phase, the SC realizes the topological spin Hall effect in a higher-order fashion: (i) The edge states emerging in the bulk band gap exhibits partial spin-momentum locking and are gapped due to the reduced spatial symmetry at the edges, (ii) The gapped edge states stabilize the topological corner states emerging in the edge band gap. The physics revealed here can be generalized to other physical systems, where the manipulation of edge and corner states with pseudospins enable unprecedented control of waves in metamaterials.

*Note added:* We notice a recent work on the higher-order quantum spin Hall effect of light [55] at the final stage of this work. The results in Figs. 3(e) and 3(f) are inspired by the results in Ref. [55].

## **Appendix**

To reveal the underlying physics of the topological phase transition in our SCs, we exploit the $k \cdot p$ theory to derive the effective Hamiltonian for the second, third, fourth and fifth acoustic bands around the $\Gamma$ point. The acoustic wave eigen-equation for the 2D SC is written as

$$-\nabla \cdot [\rho^{-1}(\mathbf{r})]\nabla \psi_{n,\mathbf{k}}(\mathbf{r}) = \omega_{n,\mathbf{k}}^2 \kappa^{-1}(\mathbf{r})\psi_{n,\mathbf{k}}(\mathbf{r}), \qquad (1)$$

where $\psi_{n,\mathbf{k}}(\mathbf{r})$ and $\omega_{n,\mathbf{k}}$ are the wavefunction (for the pressure filed) and the eigen-frequency of the acoustic Bloch states with wavevector $\mathbf{k}$ in the *n*-th band, $\kappa(\mathbf{r})$ and $\rho(\mathbf{r})$ denote the constitutive compress modulus and mass density which depend on the 2D coordinate vector $\mathbf{r}$, respectively. The acoustic Bloch wavefunctions are orthogonal, satisfying the normalization condition, $\delta_{nn'} = \int_{u.c.} \psi^*_{n,\mathbf{k}}(\mathbf{r})\kappa^{-1}(\mathbf{r})\psi_{n',\mathbf{k}}(\mathbf{r})d\mathbf{r}$ ($u.c.$ denotes the unit-cell). According to the $k \cdot p$ theory, $\psi_{n,\mathbf{k}}(\mathbf{r})$ is expanded by the Bloch wavefunctions at the $\Gamma$ point, i.e., $\psi_{n,\mathbf{k}}(\mathbf{r}) = e^{i\mathbf{k}\cdot\mathbf{r}}\sum_{n'} C_{n,n'}\psi_{n',\Gamma}(\mathbf{r})$, where $C_{n,n'}$ is the expansion coefficients. The wavevector $\mathbf{k}$ is treated as a small quantity and the Hamiltonian is expressed as the Taylor expansion of $\mathbf{k}$. Direct calculation yields the following $k \cdot p$ Hamiltonian,

$$H_{nn'} = \omega^2_{n,\Gamma}\delta_{nn'} + \mathbf{k}\cdot\mathbf{p}_{nn'} + \cdots, \tag{2}$$

with $\omega_{n,\Gamma}$ being the eigen-frequency of the $n$-th band at the $\Gamma$ point. The matrix element of $\mathbf{p}$ is given by

$$\mathbf{p}_{nn'} = \int_{u.c.} \psi^*_{n,\Gamma}(\mathbf{r})\{-i[2\rho^{-1}(\mathbf{r})\nabla + \nabla\rho^{-1}(\mathbf{r})]\}\psi_{n',\Gamma}(\mathbf{r})d\mathbf{r}. \tag{3}$$

A crucial fact is that the matrix element of $\mathbf{p}_{nn'}$ is nonzero only when the $n$ and $n'$ bands are of different parity. Therefore, the $k \cdot p$ Hamiltonian in the basis of $(p_+, d_+, p_-, d_-)^T$ can be written as (up to the linear order of $\mathbf{k}$),

$$H = \begin{pmatrix} \omega^2_p & Ak_+ & 0 & 0 \\ A^*k_- & \omega^2_d & 0 & 0 \\ 0 & 0 & \omega^2_p & Ak_- \\ 0 & 0 & A^*k_+ & \omega^2_d \end{pmatrix}, \tag{4}$$

where $k_\pm = k_x \pm ik_y$, and $A$ is the coupling coefficient. $\omega_p(\omega_d)$ refers to frequency of the doublet states $p_\pm(d_\pm)$ at the $\Gamma$ poin. We notice that above Hamiltonian is similar to the Bernevig-Hughes-Zhang model for the quantum spin Hall insulator where the "spins" are emulated by the acoustic OAM. Using this analog, when the frequency of the $p$-state is lower than that of the $d$-state, as for the case $\theta = 0°$, the 2D SC is a normal insulator. In contrast, when the frequency of the $p$-state is higher than that of the $d$-state, as for the case $\theta = 60°$, the 2D SC is an acoustic analog of the quantum spin Hall insulator.